\documentclass[aps,prb,twocolumn,superscriptaddress,showpacs]{revtex4-2}
\usepackage{amsmath,amssymb,wasysym,graphicx}
\usepackage[utf8]{inputenc}
\usepackage[T1]{fontenc}
\usepackage{xcolor}

\IfFileExists{newtxtext.sty}
{\usepackage{newtxtext,newtxmath}}
{\IfFileExists{stix.sty}
	{\usepackage{stix}}
	{\IfFileExists{mathptmx.sty}
		{\usepackage{mathptmx}}{} } }

\usepackage{textcomp}

\usepackage{bm}

\IfFileExists{siunitx.sty}{\usepackage{booktabs,siunitx}}{}

\usepackage{color}
\definecolor{LinkColor}{rgb}{0.256,0.439,0.588}
\usepackage{hyperref}
\hypersetup{
	pdfauthor={good guys},
	pdftitle={good title},
	colorlinks=true,
	citecolor=LinkColor,
	linkcolor=LinkColor,
	urlcolor=LinkColor
}

\newcommand{\beq} {\begin{equation}}
\newcommand{\eeq} {\end{equation}}
\newcommand{\bea} {\begin{eqnarray}}
\newcommand{\eea} {\end{eqnarray}}
\newcommand{\be} {\begin{equation}}
\newcommand{\ee} {\end{equation}}

\newcommand{\ket}[1]{\left|#1\right>}
\newcommand{\bra}[1]{\left<#1\right|}

\begin{document}
\title{Quantum phase transition between topologically distinct quantum critical points}

\author{Xue-Jia Yu}
\altaffiliation{The first two authors contributed equally.}
\email{xuejiayu@fzu.edu.cn}
\affiliation{Department of Physics, Fuzhou University, Fuzhou 350116, Fujian, China}
\affiliation{Fujian Key Laboratory of Quantum Information and Quantum Optics,
College of Physics and Information Engineering,
Fuzhou University, Fuzhou, Fujian 350108, China}

\author{Wei-Lin Li}
\affiliation {Key Laboratory of Atomic and Subatomic Structure and Quantum Control (Ministry of Education), Guangdong Basic Research Center of Excellence for Structure and Fundamental Interactions of Matter, School of Physics, South China Normal University, Guangzhou 510006, China} 

\affiliation {Guangdong Provincial Key Laboratory of Quantum Engineering and Quantum Materials, Guangdong-Hong Kong Joint Laboratory of Quantum Matter, South China Normal University, Guangzhou 510006, China}

\date{\today}

\begin{abstract}

By constructing an exactly solvable spin model, we investigate the critical behaviors of transverse field Ising chains interpolated with cluster interactions, which exhibit various types of topologically distinct Ising critical points. Using fidelity susceptibility as an indicator, we establish the global phase diagram, including ferromagnetic, trivial paramagnetic, and symmetry-protected topological phases. Different types of critical points exist between these phases, encompassing both topologically trivial and non-trivial Ising critical points, as well as Gaussian critical points. Importantly, we demonstrate the existence of a Lifshitz transition between these topologically distinct Ising critical points, with central charge and critical exponents determined through finite-size scaling. This work serves as a valuable reference for further research on phase transitions within the gapless quantum phase of matter.

\end{abstract}

\maketitle

\section{INTRODUCTION}
\label{sec:introduction}
The classification of phases and phase transitions is a foundational issue in condensed matter and statistical physics \cite{sachdev_2011,sachdev_2023,fradkin2013field,sondhi1997}. The traditional paradigm of phase transition relies on the Landau-Ginzberg-Wilson symmetry-breaking paradigm \cite{cardy_1996,yu2022fidelity}. However, since the 1980s, the development of topological phases of matter has received significant attention \cite{wen2017rmp,HALDANE1983464,haldane1983prl}, expanding our comprehension of quantum matter beyond the Landau paradigm \cite{xu2012unconventional,senthil2023deconfined}.A notable example is the symmetry-protected topological (SPT) phase \cite{chen2010prb,chen2013prb,chen2012symmetry}. It's worth noting that discussions of SPT phases typically focus on gapped quantum phases \cite{chen2011prb} in the past few decades. Nevertheless, there are large unexplored areas within the field of gapless quantum phases of matter, particularly in the context of gapless topological phases.

Although extensive research has been conducted on non-interacting gapless topological phases, such as Dirac or Weyl semimetals~\cite{yan2017topological,vafek2014dirac,armitage2018rmp}, there has been a notable scarcity of studies addressing strongly interacting gapless topological phases. These phases, considered as direct extensions of the SPT phase, have been discussed in the literature~\cite{scaffidi2017prx,parker2018prb,ryan2021prb,yang2023prb,wen2023prb,hidaka2022prb,li2023decorated,li2023intrinsicallypurely}. They are often referred to as gapless SPT (gSPT) or symmetry-enriched quantum critical points~\cite{ruben2021prx,yu2022prl,duque2021prb,huang2023emergent,ye2022scipost,mondal2023symmetryenriched,wang2023stability,jones2021prr,smith2022prr,yu2024universal}. These quantum phases exhibit trivial bulk properties but with anomalous boundary behavior, which closely aligns with recent investigations into the boundary criticality of both classical and quantum systems~\cite{max2022scipost,parisen2022prl,padayasi2022scipost,krishnan2023plane,sun2023extraordinarylog,sun2022prb,hu2021prl,zhu2021prb,zhang2017prl,ding2018prl,weber2018prb,jian2021scipost,xu2020prb,wuxiaochuan2020prb}. Furthermore, recent literature has proposed a method called the Pivoet Hamiltonian, which offers a systematic approach for constructing topologically distinct quantum critical points~\cite{nat2023scipost,nat2023scipost_b}, making it very convenient to study the phase transition between them. There has been a significant surge in progress towards simulating quantum phases of matter characterized by nontrivial entanglement using platforms summarized under the category of noisy intermediate-scale quantum (NISQ) technology~\cite{preskill2018quantum,sahay2023quantum}. These advancements include the simulation of exotic quantum many-body states, such as topological order, spin liquids, SPT phases, and unconventional quantum phase transition~\cite{semeghini2021probing,satzinger2021realizing,de2019observation,keesling2019quantum,song2018prl,nat2023prl,nat2023prxquantum,tantivasadakarn2022longrange,verresen2022efficiently,zhu2022nishimoris,iqbal2023topological,iqbal2023creation,chen2023realizing,yuxuejia2023pra,yangsheng2023prb,Yu2024PRL}, which have long been topics of discussion in the field of condensed matter and statistical physics.

However, the phase transition between topologically distinct quantum critical points (QCPs) or gapless phases is rarely mentioned. To address these issues, fidelity susceptibility, a concept borrowed from quantum information theory \cite{gu2010fidelity,albuquerque2010prb}, offers a remarkably simple and intuitive method for identifying QCPs. To date, fidelity susceptibility has proven effective in detecting various QCPs, including conventional symmetry-breaking QCPs \cite{zhu2018pra,sun2017pra,yu2023pre}, topological phase transitions~\cite{sun2015prb}, Anderson transitions~\cite{wei2019pra,lv2022pra,lishanzhong2023prb}, non-conformal commensurate-incommensurate transitions~\cite{yu2022fidelity}, deconfined quantum criticality~\cite{sun2019prb}, and even non-Hermitian critical points \cite{sun2022biorthogonal,tzeng2021prr,tu2022general,yu2023nonhermitian}. Nevertheless, it remains an open question whether fidelity susceptibility can effectively detect the quantum critical and scaling behaviors in gapless-gapless phase transitions, particularly those involving the transition between topologically distinct universality classes, often referred to as the "transition" of phase transitions.

In this work, we answer the series of questions outlined above by constructing an exactly solvable spin model, which is a linear combination of transverse field and cluster Ising models. Employing the Jordan-Wigner transformation, we have thoroughly examined various properties of the model, including the ground-state energy density, winding number, fidelity susceptibility, entanglement entropy, and order parameters. These investigations have allowed us to establish the global phase diagram and comprehend its critical behaviors. Moreover, we not only pay attention to the critical behaviors of the non-conformal Lifshitz transition point between topologically distinct Ising universality classes but also investigate the conformal phase transition between the SPT and paramagnetic (PM) phase.

The paper is organized as follows: Sec.~\ref{sec:model} contains the lattice model of the quantum Ising chain interpolated with cluster interaction. Section ~\ref{sec:potts} shows the global phase diagram of the model and the finite-size scaling for various physical quantities. The conclusion
is presented in Sec.~\ref{sec:con}. Additional data for our analytical and numerical calculations are provided in the Appendix.
\section{MODEL AND METHOD}%
\label{sec:model}
\subsection{Quantum Ising chain interpolated with cluster interaction}
The system under study is a quantum Ising chain interpolated with a three-body cluster interaction. The model is defined by the following Hamiltonian:
\begin{equation}
\begin{split}
\label{E1}
& H = \lambda H_{TFI}+(1-\lambda)H_{CI}, \\
& H_{TFI} = -\sum_{j=1}^{N-1}\sigma^{x}_{j}\sigma^{x}_{j+1}-h\sum_{j=1}^{N}\sigma^{z}_{j}, \\
& H_{CI} = -\sum_{j=1}^{N-1}\sigma^{x}_{j}\sigma^{x}_{j+1}+h\sum_{j=1}^{N-2}\sigma^{x}_{j}\sigma^{z}_{j+1}\sigma^{x}_{j+2}.
\end{split}
\end{equation}
Here, $\sigma_{i}^{x/y/z}$ represents the spin-$\frac{1}{2}$ Pauli matrices on each site $i$. The Hamiltonians $H_{TFI}$ and $H_{CI}$ correspond to the transverse field and the cluster Ising model, respectively. Notably, these models possess a $\mathbb{Z}_{2}$ spin-flip (generated by $P = \prod_{i}\sigma^{z}_{i}$) and a time-reversal symmetry denoted as $\mathbb{Z}^{T}_{2}$ (acting as the complex conjugation $T=K$). The $\lambda$ serves as a tuning parameter that governs the competition between two different quantum spin chains, ultimately leading to the emergence of an unconventional universality class.

Although both the spontaneous symmetry-breaking phase (e.g., ferromagnetic (FM) phase) to trivial PM or SPT phase transitions are described by Ising conformal field theory (CFT), the distinct behavior of the time-reversal symmetry towards the symmetry flux operator (also known as the disorder operator) gives rise to topologically distinct (symmetry-enriched ) QCPs or gSPT \cite{ruben2021prx,duque2021prb}. To provide a brief overview of this distinction, it's important to consider that an Ising CFT has a unique local primary field denoted as $\sigma$ with a scaling dimension $\Delta = 1/8$, as well as a unique nonlocal primary field denoted as $\mu$ with the same scaling dimension. These primary fields correspond to the order parameters of the nearby phases. For instance, $\sigma(n) \sim \sigma^{x}_{n}$ is the Ising order parameter, whereas the nonlocal operator $\mu(n)$ is the Kramers-Wannier dual disorder order parameter of the disorder symmetric phases. More precisely, $\mu(n) \sim \prod_{j=-\infty}^{n}\sigma^{z}_{j}$ in the trivial PM phase, whereas $\mu(n) \sim \prod_{j=-\infty}^{n}\sigma^{x}_{j-1}\sigma^{z}_{j}\sigma^{x}_{j+1}$ in the SPT phase. Notably, the two Ising critical lines are distinguished by the discrete invariant $T\mu T=\pm \mu$, indicating that they must be separated by a phase transition, essentially representing the "transition" of a phase transition. Indeed, as depicted in Fig.~\ref{fig1}, they converge at a multicritical Lifshitz point with the dynamical exponent $z=2$ \cite{ARDONNE2004493,fradkin2013field}. More broadly, one of the authors of this paper proposes that the conformal boundary condition can serve as a more general "topological invariant" for classifying topologically distinct quantum critical points, even in the absence of degenerate edge modes \cite{yu2022prl}.

In this work, we denote the topological non-trivial case, characterized by the property where the nonlocal disorder operator is charged as $T \mu T = -\mu$, as the "symmetry-enriched ${\rm{Ising^{*}}}$" critical point, which exhibits degenerate zero-energy edge modes even the bulk is gapless. At a fundamental level, the boundary of the symmetry-enriched ${\rm{Ising^{*}}}$ critical point spontaneously breaks the $\mathbb{Z}_{2}$ symmetry, creating an intriguing degenerate boundary fixed point that remains stable and corresponds to a fixed boundary condition. Remarkably, the finite-size splitting of this edge mode $\sim 1/N^{14}$, is parametrically faster than the finite-size bulk gap $\sim 1/N$. Therefore, the degenerate edge mode can maintain stability even if the bulk remains gapless. In the subsequent few sections, our focus remains on the case where $h=h_{c}=1.0$ to ensure that the system resides within the QCPs.

\section{PHASE DIAGRAM AND CRITICAL BEHAVIOR}
\label{sec:potts}

\subsection{Quantum phase diagram}

\begin{figure}[t]
\includegraphics[width=0.5\textwidth]{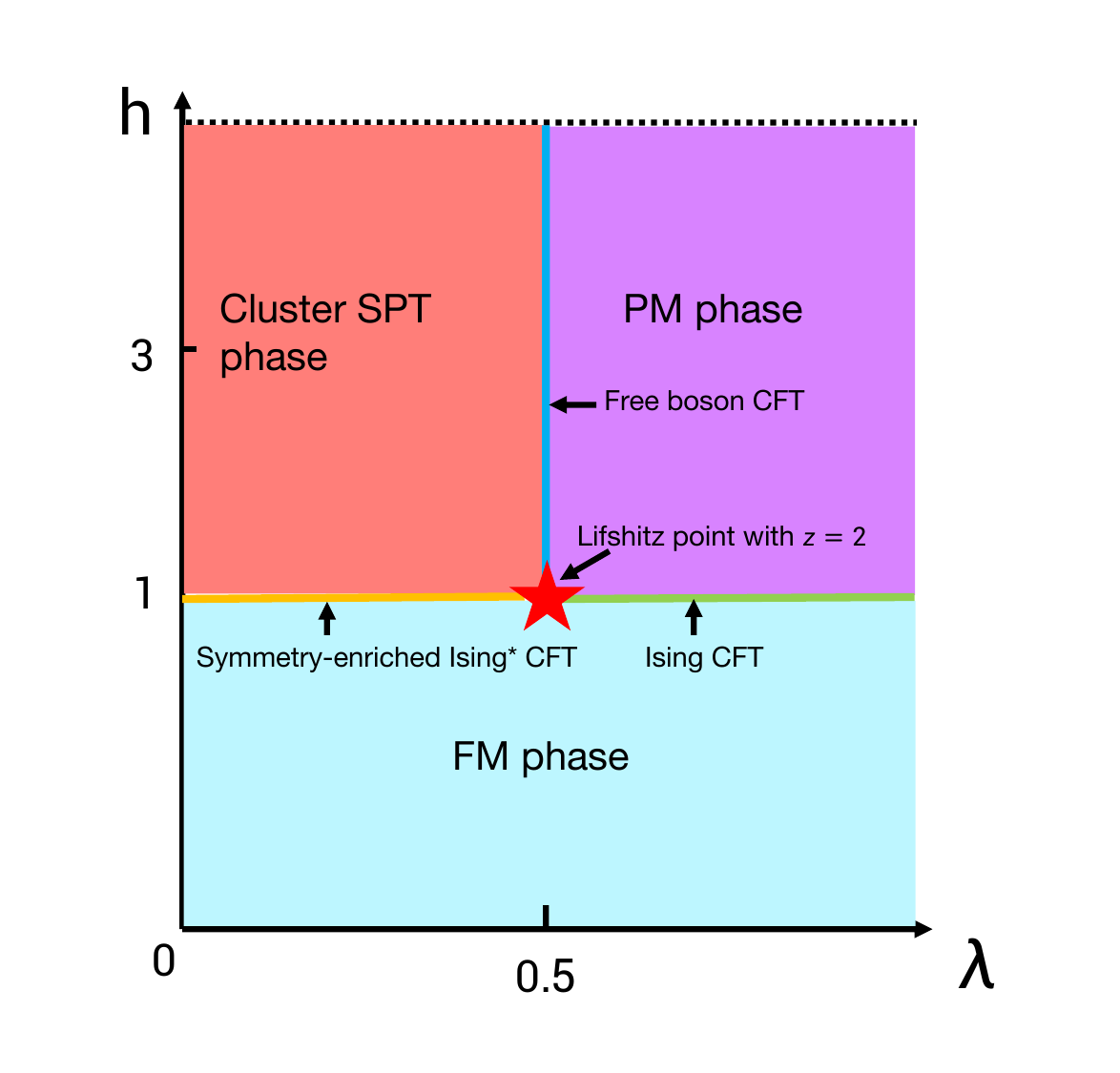}
\caption{(Color online) Schematic global phase diagram of quantum Ising model interpolated with cluster interaction in terms of tuning parameters ($\lambda,h$). The phase diagram comprises three distinct regions: the $\mathbb{Z}_{2} \times \mathbb{Z}^{T}_{2}$ cluster SPT phase (light red area), the PM phase (purple area), and the FM order phase (light blue area). When $h<1.0$, the ground state belongs to the FM order phase. When $h=1.0$, the orange (green) solid critical line represents the topological (nontrivial) trivial Ising universality class between the  FM to (cluster SPT) PM phases. For $h>1.0$, the transition from cluster SPT to PM phase (blue solid line) is described by the free boson CFT with $c=1$. The red star denotes the multicritical Lifshitz point with dynamical exponent $z=2$.}
\label{fig1}
\end{figure} 

Before delving into the analytical results, let's summarize our main findings and outline the global quantum phase diagram of the model in Eq.~(\ref{E1}). The schematic phase diagram is provided in Fig.\ref{fig1}. The tuning parameters ($h$,$\lambda$)  drive the system toward different phases, including FM, trivial PM, and $\mathbb{Z}_{2}\times \mathbb{Z}^{T}_{2}$ SPT phases \cite{Son_2011,smacchia2011pra,ruben2017prb}. The latter is sometimes referred to as the cluster or Haldane SPT phase. Furthermore, there exist rich QPTs between these quantum phases, including the 1+1D conformal (topologically trivial) Ising universality class (green solid line), symmetry-enriched (topologically nontrivial) ${\rm{Ising^{*}}}$ universality class (orange solid line), Gaussian universality class (blue solid line), and nonconformal Lifshitz criticality (red star).

To elaborate, using the integrability of the model, Eq.~(\ref{E1}) can be reformulated as a free fermion model \cite{ding2019pre,yu2022pra} through the Jordan-Wigner transformation:
\begin{equation}
\begin{split}
\label{EJW}
&\sigma^{x}_{i} = \prod_{j<i}(1-2c^{\dagger}_{j}c_{j})(c_{i}+c^{\dagger}_{i}),\\
&\sigma^{y}_{i} = -\mathrm{i} \prod_{j<i}(1-2c^{\dagger}_{j}c_{j})(c_{i}-c^{\dagger}_{i}),\\
&\sigma^{z}_{i} = 1- 2 c^{\dagger}_{i}c_{i}.
\end{split}
\end{equation}
After applying the Fourier transformation $c_{k}=\frac{1}{\sqrt{N}}\sum_{j=1}^{N}e^{ikj}c_{j}$, where $k=2\pi m/N$ and $m$ ranges from $-(N-1)/2$ to $(N-1)/2$, we obtain the following free Hamiltonian:

\begin{equation}
\begin{split}
\label{E3}
& H(h,\lambda) = 2\sum_{k>0}[\mathrm{i}y_{k}(c^{\dagger}_{k}c^{\dagger}_{-k}+c_{k}c_{-k}) \\
&+ z_{k}(c^{\dagger}_{k}c_{k}+c^{\dagger}_{-k}c_{-k}-1)] + {\rm{const}},
\end{split}
\end{equation}
where $y_{k} = -{\rm{sin}}(k) + (1-\lambda)h {\rm{sin}}(2k)$ and $z_{k} = \lambda h -{\rm{cos}}(k) + h(1-\lambda){\rm{cos}}(2k)$. Subsequently, the Hamiltonian takes on a bilinear form and can be diagonalized using the Bogoliubov transformation:

\begin{equation}
\begin{split}
\label{E4}
& b_{k} = {\rm{cos}}(\frac{\theta_{k}}{2})c_{k}-\mathrm{i}{\rm{sin}}(\frac{\theta_{k}}{2})c^{\dagger}_{-k},\\
& b^{\dagger}_{k} = {\rm{cos}}(\frac{\theta_{k}}{2})c^{\dagger}_{k}+\mathrm{i}{\rm{sin}}(\frac{\theta_{k}}{2})c_{-k},\\
& H(h,\lambda) = \sum_{k>0}\epsilon_{k}(b^{\dagger}_{k}b_{k}-\frac{1}{2}),
\end{split}
\end{equation}
where $b_{k}$ ($b_{k}^{\dagger}$) is the Bogoliubov quasiparticle annihilation (creation) operator, $\epsilon_{k}=4\sqrt{y_{k}^{2}+z_{k}^{2}}$, ${\rm{tan}}(\theta_{k})=-\frac{y_{k}}{z_{k}}$, and the ground state is given by: $|G \rangle = \prod_{k>0}[{\rm{cos}}(\frac{\theta_{k}}{2})+\mathrm{i}{\rm{sin}}(\frac{\theta_{k}}{2})c^{\dagger}_{k}c^{\dagger}_{-k}]|{\rm{Vac}}\rangle$ ($|{\rm{Vac}} \rangle$ is the vacuum state of $c$ fermion).

\begin{figure}[t]
\includegraphics[width=0.45\textwidth]{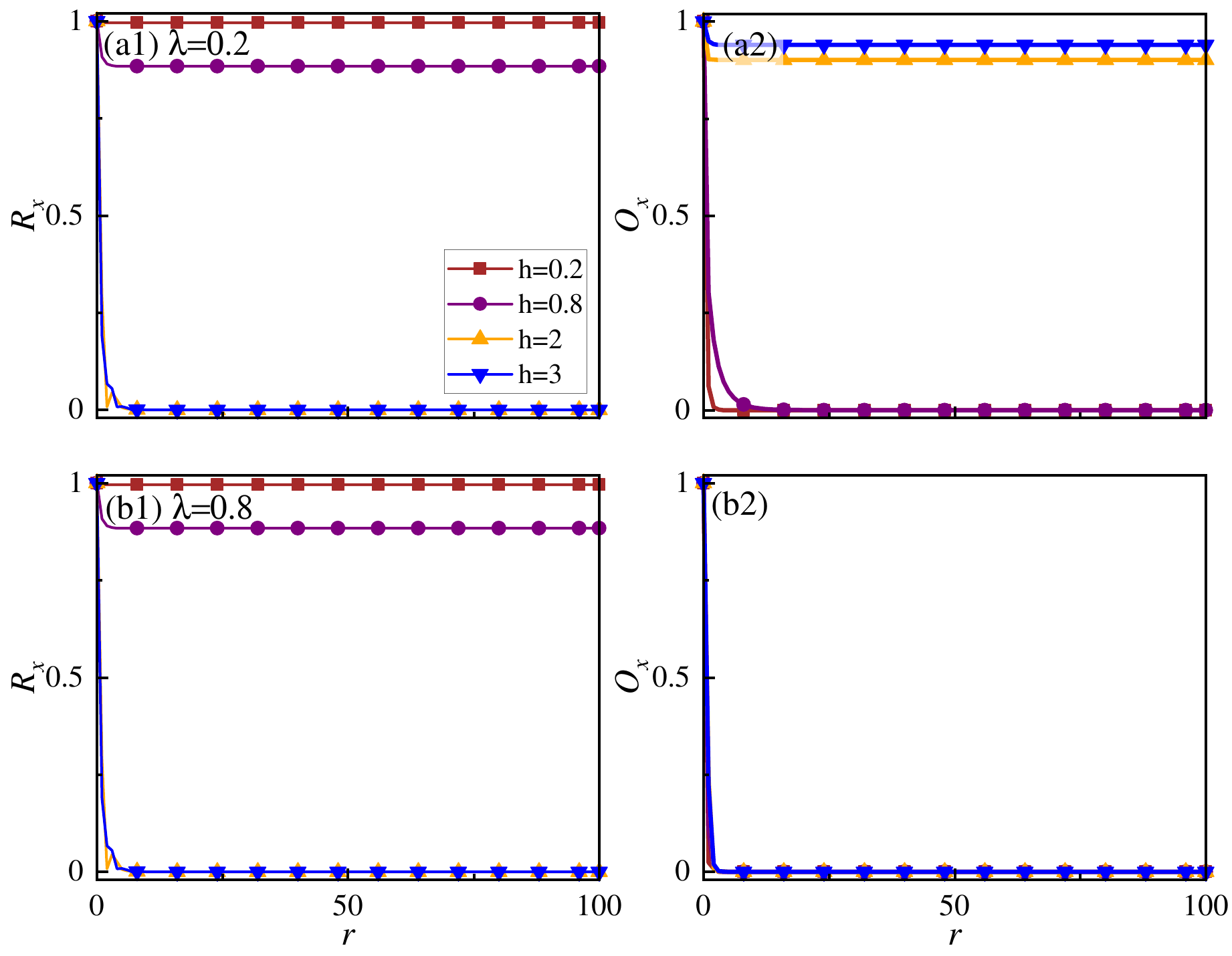}
\caption{(Color online) The spin correlation $|R_{x}(r)|$ is plotted as a function of distance $r$ for different $h$ with $\lambda=0.2$ (a1) and $0.8$ (b1), while the string order parameter $|O_{x}(r)|$ is depicted as a function of $r$ for $\lambda=0.2$ (a2) and $0.8$ (b2). Notably, when $h<1.0$, regardless of whether $\lambda$ is greater or less than $0.5$, the FM spin correlation exhibits long-range order, indicating that the ground state features FM order. Conversely, when $h>1.0$ and $\lambda < 0.5$ ($\lambda=0.2$), the string order parameter displays long-range order, suggesting the presence of a cluster SPT phase. Finally, in the scenario where $h>1.0$ and $\lambda > 0.5$ ($\lambda=0.8$), both the string order parameter and FM correlation exhibit short-range behaviors, indicative of a trivial PM phase.}
\label{fig2}
\end{figure} 

Before delving into phase transitions, let's explore the possible phases that appear in a phase diagram. As a preliminary step, we examine some limiting cases: When $\lambda=0.0$, the model is simplified to a cluster Ising model. By adjusting the parameter $h$, the model can achieve a phase transition from the FM (small $h$) to the cluster SPT phase (large $h$). Conversely, when $\lambda=1.0$, the model is reduced to the usual transverse field Ising model. At this time, adjusting the parameter $h$ can achieve the phase transition from the FM (small $h$) to the trivial PM phase (large $h$). In general cases ($\lambda,h$), to identify the possible quantum phase of matter, we calculated the FM correlation function (order parameter) and string order parameter:
\begin{equation}
\begin{split}
\label{order_parameter}
& R_{x}(r)=\frac{1}{N} \sum_{i=1}^{N} \langle \sigma^{x}_{i}\sigma^{x}_{i+r} \rangle,\\
& O_{x} = \lim_{N \rightarrow \infty} \langle \sigma^{x}_{1} \sigma^{y}_{2} (\prod_{k=3}^{N-2}\sigma^{z}_{k})\sigma^{y}_{N-1}\sigma^{x}_{N} \rangle.
\end{split}
\end{equation}
As depicted in Fig.~\ref{fig2}(a1) and (a2), we observe that when $\lambda < 0.5 $ ($\lambda=0.2$) and $h<1.0$ ($h=0.2,0.8$), the FM or string order parameter remains constant or tends to zero in the long-distance limit, suggesting the existence of FM long-range order in this region~\cite{sachdev_2011}. Conversely, when $\lambda < 0.5$ ($\lambda = 0.2$) and $h>1.0 (h=2.0,3.0)$, the string order parameter or FM spin correlation function becomes constant or zero in the long-distance limit, indicating that the system resides in the cluster SPT phase in such a parameter region~\cite{ruben2017prb}.

Similarly, as illustrated in Fig.~\ref{fig2}(b1) and (b2), when $\lambda > 0.5 $ ($\lambda=0.8$) and $h<1.0$ ($h=0.2,0.8$), the FM or string order parameter remains constant or zero in the long-distance limit, implying FM long-range order dominate in this region. However, when $\lambda > 0.5 $ ($\lambda=0.8$) and $h>1.0$ ($h=2.0,3.0$), both the string order parameter and FM correlation tends to zero under long-distance limits, indicating that the system is in the trivial PM phase in such a region.

\subsection{"Transition" of phase transition}
After delineating all the quantum phases in the phase diagram, we shift our focus to the more intriguing QPTs between these phases. While traditional discussions primarily concentrate on the phase transitions between gapped phases, it's entirely plausible that there exists an unconventional QPTs between different gapless topological phases \cite{shi2022prb}. For simplicity, our attention is drawn to the intriguing "transition" between topologically distinct critical points. For our purposes, we set $h=1.0$ in the model Eq~(\ref{E1}), and by manipulating the parameter $\lambda$, we first consider two tractable cases:

1) When $\lambda=0.0$: the model corresponds to a critical cluster Ising chain, thereby realizing the symmetry-enriched ${\rm{Ising^{*}}}$ universality class~\cite{ruben2021prx,duque2021prb}.

2) When $\lambda=1.0$: the model transforms into a usual critical Ising chain, belonging to the 1+1D (topological trivial) Ising universality class.

These two (topologically) distinct Ising universality classes correspond to different conformal boundary conditions~\cite{yu2022prl,yu2024universal}, and it is infeasible to smoothly connect them without either breaking the symmetry or encountering a multicritical point. Therefore, akin to the unconventional phase transition between gapped topological phases~\cite{ruben2017prb}, there may exist an unconventional QPT between topologically distinct quantum critical points or critical phases, which constitutes a largely unexplored area in statistical and condensed matter physics. In the following subsections, we provide evidence of unconventional phase transitions from different perspectives.

\begin{figure}[t]
\includegraphics[width=0.45\textwidth]{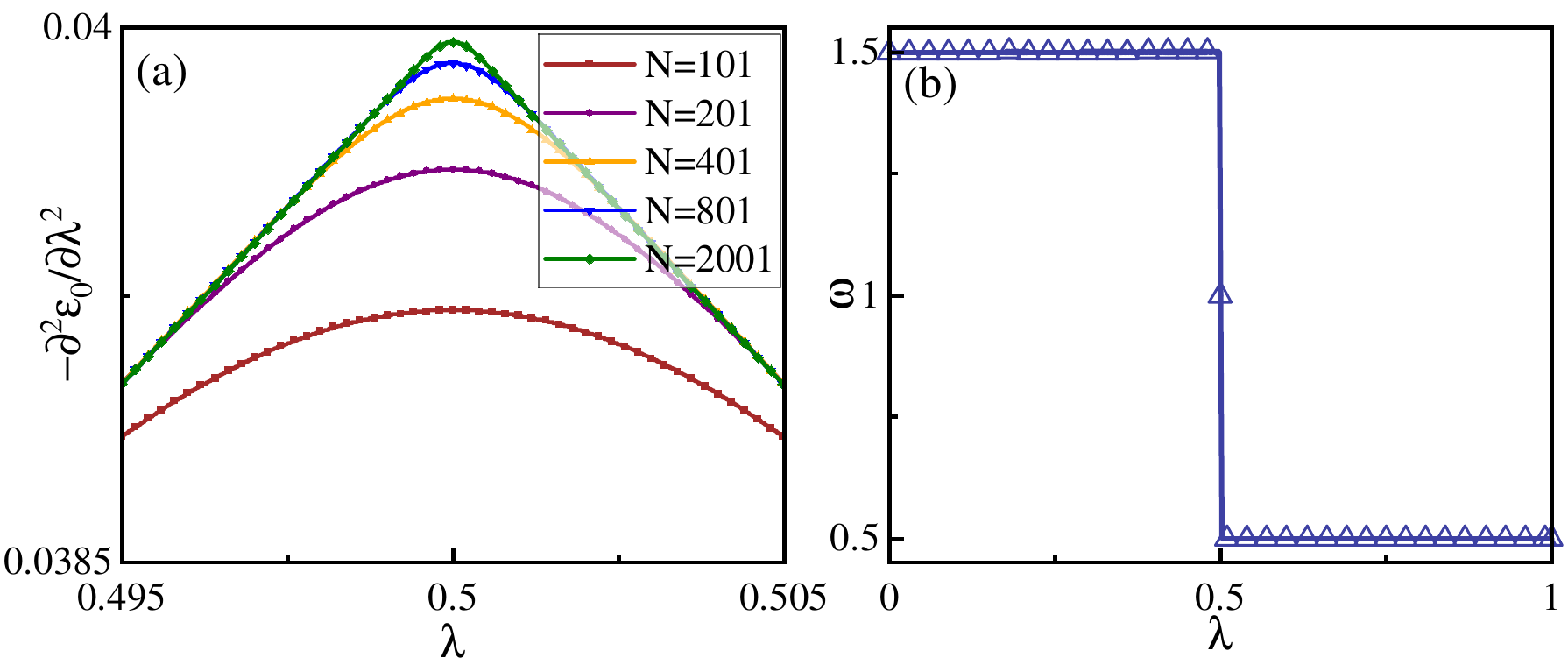}
\caption{(Color online) (a) The second derivative of ground-state energy density
$-\frac{\partial^{2} \varepsilon_{0}}{\partial \lambda^{2}}$ with respect to $\lambda$ for $h=1.0$. (b) The winding number as a function of $\lambda $ with $h=1.0$ for $N=2000$.}
\label{fig3}
\end{figure} 

\subsubsection{Ground state energy density and its second-order derivative}

According to Eq.~(\ref{E4}), the ground state energy density of the model is expressed as:

\begin{equation}
\label{E5}
\varepsilon_{0} = -\sum_{k>0}\frac{\epsilon_{k}}{2N} = -\frac{2}{N}\sum_{k>0}\sqrt{y_{k}^{2}+z_{k}^{2}}=-\frac{1}{\pi}\int_{0}^{\pi}\sqrt{y_{k}^{2}+z_{k}^{2}}dk.
\end{equation}

Using this equation, we can numerically calculate both $\varepsilon_{0}$ and its second-order derivative $-\frac{\partial^{2}\varepsilon_{0}}{\partial \lambda^{2}}$.
As illustrated in Fig.~\ref{fig3}(a), we observe that the second derivative of the ground state energy density with respect to $\lambda$ becomes sharper at $\lambda=0.5$ as the system size increases. This suggests that $\lambda=0.5$ serves as a critical point between two distinct Ising universality classes.

\subsubsection{Winding number}
Following Ref.~\cite{zhang2015prl}, we can express Eq.~(\ref{E4}) as:

\begin{equation}
\label{E6}
H(h,\lambda) = 4\sum_{k>0}\vec{h_{k}}\cdot \vec{s_{k}}.
\end{equation}
Here, $\vec{h_{k}} = (0,y_{k},z_{k}) $ and the pseudospin $\vec{s_{k}} = [(c^{\dagger}_{-k}c_{k}-c^{\dagger}_{k}c_{-k})/2,i(c^{\dagger}_{k}c^{\dagger}_{-k}+c_{k}c_{-k})/2,(c^{\dagger}_{k}c_{k}+c^{\dagger}_{-k}c_{-k}-1)/2]$. These pseudospin operators satisfy the $SU(2)$ algebra.

The winding number is defined in the parameter space $(y,z)$ as:

\begin{equation}
\label{E7}
\omega = \frac{1}{2\pi} \int_{c} \frac{1}{h^{2}}(zdy-ydz),
\end{equation}

here, $c$ represents the loop in the $(y,z)$ space as $k$ varies from $0$ to $2\pi$. $\omega$ serves as a means to distinguish between different topological phases, which possess a different winding number.

As depicted in Fig.~\ref{fig3}(b), we observe a jump in the winding number at $\lambda=0.5$, signifying a topological phase transition at this specific point. Additionally, we have determined that the winding number takes on fractional values (0.5 and 1.5) at the usual topological trivial and nontrivial Ising critical points, respectively, aligning with findings in previous literature \cite{duque2021prb,verresen2020topology}.

\subsubsection{Entanglement entropy and central charge}
Quantum entanglement serves as a powerful tool in describing QPTs, with entanglement entropy being the most commonly used quantity for this purpose. In a quantum many-body system, entanglement entropy characterizes the QPT induced by a tuning parameter by properly extracting it from the ground state wavefunction $\ket{\psi_{0}}$. Typically, the Hamiltonian is divided into two subsystems $A$ and $B$, and the reduced density matrix for subsystem $A$ is computed by tracing over the degrees of freedom of subsystem $B$, given by:
\begin{equation}
\label{E7}
\rho_{A} = {\rm{Tr}}_{B}(\ket{\psi_{0}}\bra{\psi_{0}}).
\end{equation}
The entanglement entropy, measuring the entanglement between parts $A$ and $B$, is then expressed as:
\begin{equation}
\label{E8}
S_{A} = -{\rm{Tr}}(\rho_{A}{\rm{log}}(\rho_{A})),
\end{equation}
which is evaluated in terms of the eigenvalues of $\rho_{A}$. For a one-dimensional short-range interacting system with periodic boundary conditions, CFT suggests that the entanglement entropy for subsystem $A$ with size $l$ follows the finite-size scaling behavior~\cite{Calabrese_2009,li2023pra}
\begin{equation}
\label{E9}
S_{l} \sim \frac{c}{3}\text{ln}(\frac{N}{\pi}{\rm{sin}}(\frac{\pi l}{N})) + S^{\prime},
\end{equation}
where $c$ is the central charge, which varies for different universality classes, and $S^{\prime}$ is a non-universal constant.

Back to our case, since the Hamiltonian in Eq.~(\ref{E1}) is quadratic and exactly solvable, the ground state $\ket{\psi_{0}}$ is a BCS-type state, and its correlation matrix $D_{ij} = \bra{\psi_{0}}c^{\dagger}_{i}c_{j}\ket{\psi_{0}}$ can be efficiently analytical calculated. Therefore, the entanglement entropy $S_{A}$ between subsystems $A$ and $B$ can be easily obtained as~\cite{vidal2003prl,li2023emergent}
\begin{equation}
\label{E10}
S_{A} = -{\rm{Tr}}[D_{A}{\rm{log}}(D_{A})+(1-D_{A}){\rm{log}}(1-D_{A})],
\end{equation}
where $D_{A}$ is the correlation matrix for subsystem $A$ =$\{1,2,...,l\}$.

For $h=1.0$, in addition to the multicritical point $\lambda = 0.5$, there exists a critical line belonging to the Ising universality class (see Appendix~\ref{sec:A5} for details). Consequently, the entanglement entropy of the system adheres to the scaling law of CFT, with a central charge of $c=0.5$ (see Appendix \ref{sec:A1} for details). The numerical findings in the preceding sections suggest that when $\lambda =0.5$, representing the "phase transition" transition, the critical point belongs to Lifshitz criticality with a dynamical exponent $z = 2$ (see next section). This implies that the phase transition deviates from CFT descriptions and does not conform to the entanglement entropy scaling law mentioned earlier. However, recent studies~\cite{wang_scipost_2022,wang2023prb} have illustrated that Lifshitz transitions with $z=2$ exhibit anomalous entanglement entropy scaling behavior and universal finite-size amplitudes, thus offering a promising avenue for future research on entanglement concerning the phase transition between topologically distinct critical points or phases.

\subsubsection{Finite-size scaling and critical exponents}
To date, we have established numerically the existence of a Lifshitz multicritical point between topologically distinct Ising universality classes. This discovery naturally leads to inquiries about the scaling and critical exponents at this multicritical point. In this work, we obtained the critical exponents through the finite-size scaling of fidelity susceptibility.

The concept of fidelity susceptibility pertains to a system undergoing a continuous phase transition from an ordered to a disordered phase upon tuning the parameter $\lambda$ to a critical value $\lambda_{c}$. At this point, the structure of the ground state wave function changes significantly. The quantum ground-state fidelity $F(\lambda,\lambda + \delta \lambda)$ quantifies the overlapping amplitude between the ground state wave function at external field $\lambda$ and $\lambda+\delta \lambda$~\cite{gu2010fidelity,Gu2009pre,Gu_2009,You_2015}. Near $\lambda_{c}$, $F(\lambda_{c},\lambda_{c} + \delta \lambda)\sim 0$, indicating a drastic change in the ground state. Then, the fidelity susceptibility, defined as the leading term in the fidelity:

\begin{equation}
\begin{split}
\label{E2}
\chi_{F}(\lambda)=\lim_{\delta \lambda \rightarrow 0}\frac{2(1-F(\lambda,\lambda+\delta \lambda))}{(\delta \lambda)^{2}} = \frac{1}{4}\sum_{k>0}\left(\frac{\partial\theta_{k}(\lambda)}{\partial\lambda}\right)^2.
\end{split}
\end{equation}

\begin{figure}[t]
\includegraphics[width=0.45\textwidth]{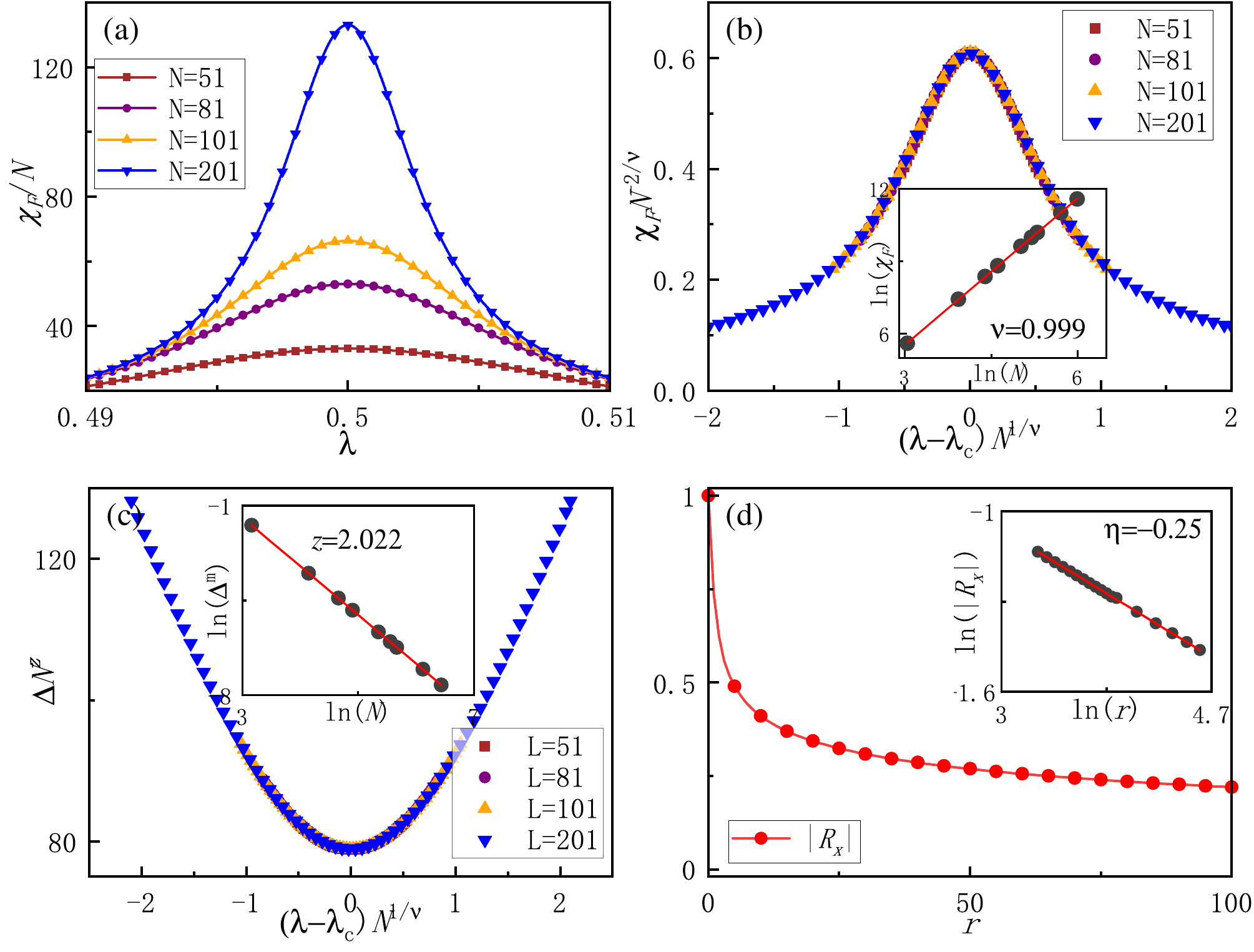}
\caption{(Color online) (a) The finite-size scaling analysis of the fidelity susceptibility per site $\chi_{N}$ for $h=1.0$. The fidelity susceptibility per site
shows a sharp peak near the transition point. (b) Data collapse of
the fidelity susceptibility per site $\chi_{N}$ and $\lambda$ with $\nu=1.0$ for various system sizes. The inset shows the log-log plot
of the fidelity susceptibility against the system size at the critical
point, and the correlation length critical exponent $
\nu=1.0$ can be inferred from the slope of the fitted straight line. (c) Data collapse of the rescaled energy gap and $\lambda$ with $z=2$,$\nu=1.0$, and $\lambda_{c} =0.5$ for
the largest four system sizes. The inset displays the log-log plot of
the energy gap $\Delta$ versus the system size $N$ at the critical
point $\lambda_{c}$, and the fitted straight line has a slope whose absolute value equals to the dynamical critical exponent $z$. (d) The variation modes FM spin correlation function $|R_{x}(r)|$ at the critical point $\lambda_{c}=0.5$ for $h=1.0$, the insets plot the curves at critical point $\lambda=0.5$ in log-log coordinates and show the slope $\eta=1/4$ of the lines.  The inset also shows the curves featuring power-law decay in ln-ln coordinates.}
\label{fig4}
\end{figure} 

For a continuous QPT in a finite system size $N$, the fidelity susceptibility $\chi_{F}(\lambda)$ exhibits a peak at a critical point, and show the finite-size scaling behaviors follow\cite{gu2010fidelity,sun2019prb,sun2015prb}
\begin{equation}
\begin{split}
\label{fs_scaling}
N^{-d}\chi_{F}(\lambda) = N^{(2/\nu)-d}f_{\chi_{F}}(N^{1/\nu}|\lambda-\lambda_{c}|),
\end{split}
\end{equation}
where $\nu$ is the critical exponent of the correlation length. $z$ is the dynamic exponent, $d$ is the spatial dimension of the system, and $f_{\chi_{F}}$ is an unknown scaling function. It's important to note that in practice, the critical exponent $\nu$ is usually extracted from fidelity susceptibility per site, $\chi_{N}(f) = \chi_{F}(\lambda)/L^{d}$.

To further investigate whether the phase transitions are described by  CFT, we calculate the energy gap $\Delta$, defined as the energy difference between the first excited state and the ground state energy. For continuous phase transitions, the energy gap is expected to vanish following $\Delta\sim\lvert{\lambda-\lambda_{c}}\rvert^{z\nu}$ as $\lambda$ approaches $\lambda_{c}$~\cite{sachdev_2011}. Combined with the divergence of the correlation length following the form $\xi\sim\lvert{\lambda-\lambda_{c}}\rvert^{-\nu}$, we obtain the scaling relation, $\Delta\sim\xi^{-z}$. Since the correlation length at the critical point of a finite system can be characterized by the lattice length $N$, the finite-size scaling form, $\Delta(\lambda_{c}, N)\propto N^{-z}$, can be finally derived. In addition, the energy gap also exhibits a similar functional form to the fidelity susceptibility~\cite{yu2022fidelity} 
\begin{equation}
\label{eq:ge_collapse}
\Delta(N) = N^{-z}\mathcal{F}_{\Delta}\big[N^{1/\nu}(\lambda-\lambda_{c})\big]\,,
\end{equation}
where $\mathcal{F}_{\Delta}$ is another scaling function associated with $\Delta$. Therefore, we can determine the dynamical exponent $z$ by performing finite-size scaling on the energy gap. 

Furthermore, we obtain another critical exponent known as the anomalous exponent, which can be extrapolated through finite-size scaling for the FM spin correlation function at the critical point:
\begin{equation}
\label{sca_eta}
|R_{x}(\lambda=\lambda_{c},r)| \sim \frac{1}{r^{\eta}},
\end{equation}
where $\eta$ is the anomalous exponent characterizing the critical universality class.

The numerical results are presented in Fig.~\ref{fig4}. Specifically, we observe a distinct peak in fidelity susceptibility at $\lambda=0.5$, which becomes more pronounced with increasing system size, as depicted in Fig. \ref{fig4}(a). This observation suggests the presence of a phase transition between two distinct Ising universality classes, consistent with the numerical results in previous sections. As shown in Fig. \ref{fig4}(b), by employing the scaling formulas of fidelity susceptibility and energy gap (Eq. (\ref{fs_scaling}), (\ref{eq:ge_collapse})), we can deduce the corresponding correlation length exponent $\nu$ and dynamical exponent $z$ (see Appendix \ref{sec:A} for analytical calculation details and Appendix \ref{sec:A5} for additional $\lambda$ values). These findings indicate the existence of non-conformal Lifshitz multicritical points between topologically distinct Ising critical points. Additionally, for a more comprehensive analysis of the critical behavior at Lifshitz points, we numerically calculated the scaling behavior of the FM spin correlation function (Eq. (\ref{order_parameter})) at the critical point, as shown in Fig. \ref{fig4}(d). The results demonstrate that the anomalous exponent is $1/4$, complementing the correlation length exponent and dynamical exponent in characterizing Lifshitz criticality.

\subsection{Topological phase transition for large $h$ limit}
For large values of $h$, the Ising interaction term becomes negligible, resulting in the simplified Hamiltonian:
\begin{equation}
\label{ETPT}
H^{\prime} = -\lambda \sum_{j=1}^{N}\sigma^{z}_{j}+(1-\lambda)\sum_{j=1}^{N-2}\sigma^{x}_{j}\sigma^{z}_{j+1}\sigma^{x}_{j+2}.
\end{equation}
When $\lambda=0.0$, the ground state corresponds to the cluster SPT phase protected by $\mathbb{Z}_{2} \times \mathbb{Z}^{T}_{2}$ symmetry. Conversely, when $\lambda=1.0$, the transverse field term dominates, and the ground state resides in the trivial PM phase. Consequently, a topological phase transition is expected between these two distinct ground states~\cite{ruben2017prb}. To systematically investigate this phase transition, similar to the previous section, we utilized the Jordan-Wigner transformation to solve the model and calculated various physical quantities such as the second derivative of the ground state energy density, winding number, entanglement entropy, and fidelity susceptibility. Additionally, we determined the central charge and critical exponent through finite-size scaling (see Appendix~\ref{sec:AppD} for details).

The numerical results reveal that the topological phase transition from the cluster SPT to the trivial PM phase is described by the free boson CFT with critical exponents $\nu=1.0$ and $z=1.0$\cite{francesco2012conformal,ginsparg1988applied,ruben2017prb}. More generally, when $h$ is finite and greater than $1.0$, this universality class of phase transitions remains stable (see Appendix~\ref{sec:AppD} for details). In other words, a critical line (blue solid line in Fig.~\ref{fig1}) characterized by the free boson CFT with $c=1$ exists in the global phase diagram.

For the sake of completeness and comparison, we briefly discuss the properties of the model around $h<1.0$. This line differs from the $h > 1.0$, we find that regardless of how $\lambda$ is tuned, our results (see Appendix~\ref{sec:AppC} for details) show that the system always exhibits FM long-range order.

\section{CONCLUSION AND OUTLOOK}
\label{sec:con}
To summarize, we investigate the phase transition between the topologically distinct QCPs, i.e., a transition of the phase transition. Using fidelity susceptibility as a diagnostic, we obtain a global phase diagram for the Hamiltonian, which interpolates between the transverse field and cluster Ising model. For $h=1.0$, by tuning the parameter $\lambda$, we observe that fidelity susceptibility detects the multicritical Lifshitz point characterized by $z=2$ and $\nu=1.0$ between different Ising universality classes. Furthermore, as a by-product, for $h>1.0$, fidelity susceptibility also identifies the phase transition between the cluster SPT and PM phases, described by $c= 1$ free boson CFT. However, for $h<1.0$, no phase transition occurs, and the ground state maintains FM order phase. Future intriguing questions involve exploring the critical behavior between topologically distinct critical points in higher dimensions and within different symmetry groups (e.g., $\mathbb{Z}_{3}$, $U(1)$, among others), as well as constructing finite-temperature phase diagrams~\cite{choi2023finite}. Our work could shed new light on the phase transition between the gapless quantum phase of matter.

\begin{acknowledgments}
We thank Shan-Zhong Li, and Zheng-Xin Guo for helpful discussions. X.-J.Yu acknowledges support from the start-up grant XRC-23102 of Fuzhou University.

\end{acknowledgments}

\bibliography{ref}

\newpage
\onecolumngrid

\appendix

\section{ANALYTICAL CALCULATION DETAILS FOR THE DYNAMICAL EXPONENT}
\label{sec:A}
In this section, we derive the expression for the single-particle fermionic excitation energy at the special momentum point $k=0$, denoted as $\epsilon_{k\sim 0}$, at the multicritical point ($h=1.0$, $\lambda=0.5$). This derivation relies on the exact solvable fermionic Hamiltonian given in Eq.~(\ref{E3}), which is obtained through the Jordan-Wigner and Bogoliubov transformation:

\begin{equation}
\begin{split}
\label{A1}
& \epsilon_{k}=\sqrt{(-{\rm{sin}}k + \frac{1}{2})({\rm{sin}}2k)^{2}+(\frac{1}{2}-{\rm{cos}}k+\frac{1}{2}{\rm{cos}}2k)^{2}},\\
& = \sqrt{({\rm{sin}}k{\rm{cos}}k-{\rm{sin}}k)^{2}+({\rm{cos}}^{2}k-{\rm{cos}}k)^{2}},\\
& = |1-{\rm{cos}}k| .
\end{split}
\end{equation}

At the low-energy momentum point $k=0$, we observe that $\epsilon_{k\sim 0 } \sim k^{2}$, indicating a dynamical exponent $z=2$ at the multicritical point. This suggests that phase transitions between topologically distinct Ising critical points cannot be described by CFT.

\section{ENTANGLEMENT ENTROPY FOR DIFFERENT $\lambda$ AT $h=1.0$}
\label{sec:A1}

In this section, we present additional data on the entanglement entropy corresponding to different values of $\lambda$ at $h=1.0$.

\begin{figure}[t]
\includegraphics[width=0.45\textwidth]{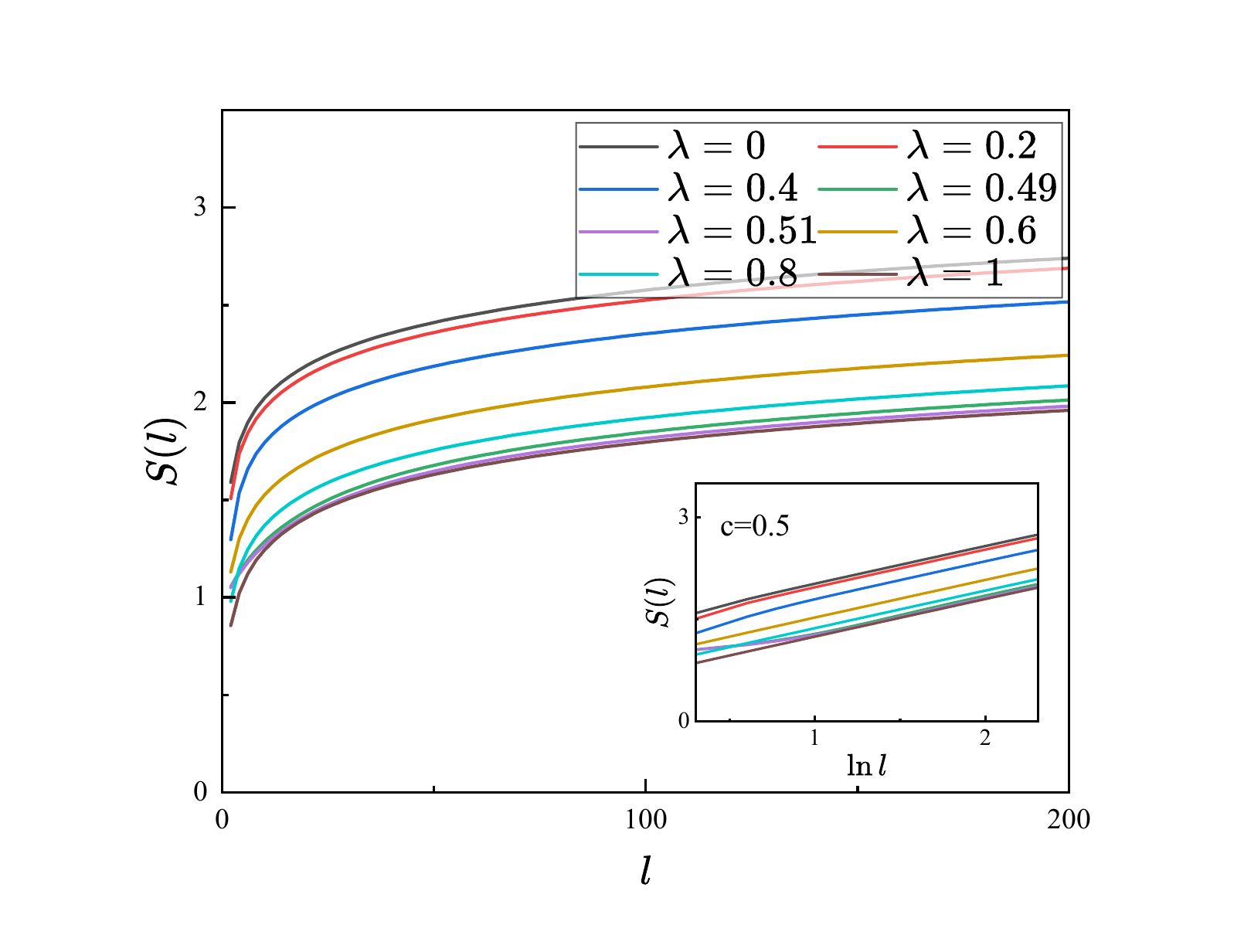}
\caption{(Color online) The entanglement entropy is plotted as a function of subsystem size $l$ for various $\lambda$ values at $h=1.0$. The central charges obtained through the fitting of entanglement entropy consistently equal 0.5, which belongs to the 1+1D Ising universality class.}
\label{sm1}
\end{figure}

Similar to the main text, we illustrate the entanglement entropy $S(l)$ as a function of subsystem sizes $l$ for $\lambda=0.0, 0.2, 0.49, 0.51, 0.8, 1.0$, and $h=1.0$ in the main panel of Fig.~\ref{sm1}. Furthermore, to determine the central charge, we provide the finite-size scaling of entanglement entropy as a function of subsystem sizes ${\rm{ln}}l$ for $\lambda = 0.0, 0.2, 0.49, 0.51, 0.8, 1.0$ in the inset of Fig. \ref{sm1}. It is apparent that the central charge at the conformal critical point consistently equals $0.5$, indicating its classification within the Ising universality class. However, as noted in the main text, the multicritical point ($\lambda=0.5$) belongs to Lifshitz criticality with a dynamical exponent $z=2$. This suggests that the phase transition is not described by CFT and exhibits anomalous entanglement entropy scaling behavior and universal finite-size amplitudes~\cite{wang_scipost_2022}.

\section{FIDELITY SUSCEPTIBILITY FOR VARUOUS $\lambda$}
\label{sec:A5}

In this section, we provide additional data on finite-size scaling for fidelity susceptibility as a function of $h$ for different $\lambda$ to ascertain the location of the Ising critical line.

\begin{figure}[t]
\includegraphics[width=0.45\textwidth]{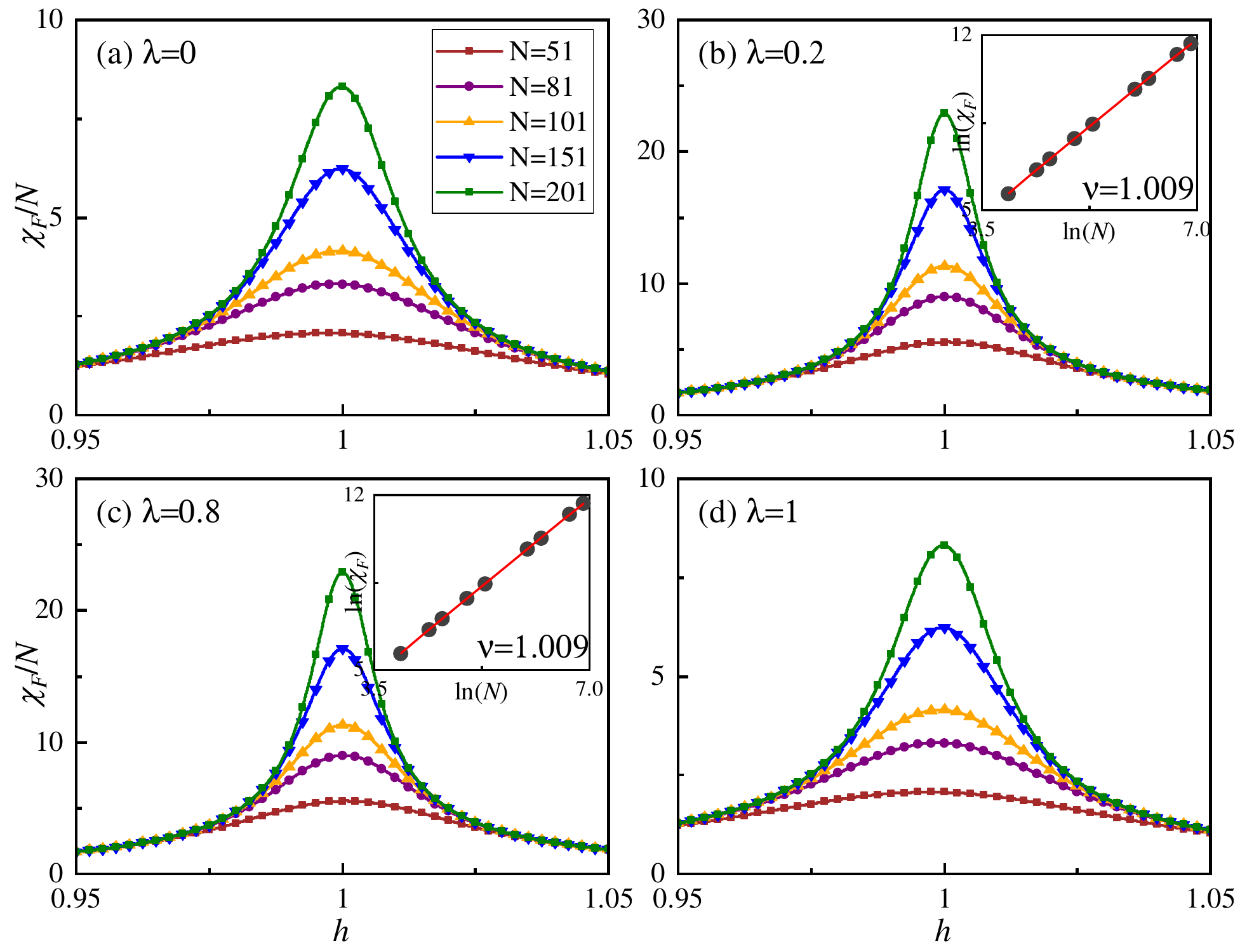}
\caption{(Color online) The finite-size scaling analysis of the fidelity susceptibility per site $\chi_{N}$ is presented for $\lambda=0.0$ (a), $0.2$ (b), $0.8$ (c), and $1.0$ (d). Notably, the fidelity susceptibility per site exhibits a sharp peak near the transition point. The insets (b) and (c) depict the log-log plot of the fidelity susceptibility against the system size at the critical point, from which the correlation length critical exponent $\nu=1.0$ can be inferred based on the slope of the fitted straight line.}
\label{sm5}
\end{figure}

As shown in Fig.~\ref{sm5}, the fidelity susceptibility demonstrates a distinct peak at $h=1.0$ across various $\lambda$ values, including $\lambda=0.0, 0.2, 0.8, 1.0$. Moreover, the sharpness of the peak increases with the system size, indicating the presence of a phase transition. Utilizing the scaling relation in Eq.~(\ref{fs_scaling}), we ascertain the correlation length exponent $\nu=1.0$ for the Ising critical point for both $\lambda=0.2$ and $\lambda=0.8$. This is illustrated in the inset of Fig.~\ref{sm5} (b) and (c). Consequently, we deduce that $h=1.0$ (excluding $\lambda=0.5$) represents an Ising critical line with a correlation length exponent $\nu=1.0$.

\section{FM LONG-RANGE ORDER FOR $h<1.0$}
\label{sec:AppC}
\begin{figure}[t]
\includegraphics[width=0.45\textwidth]{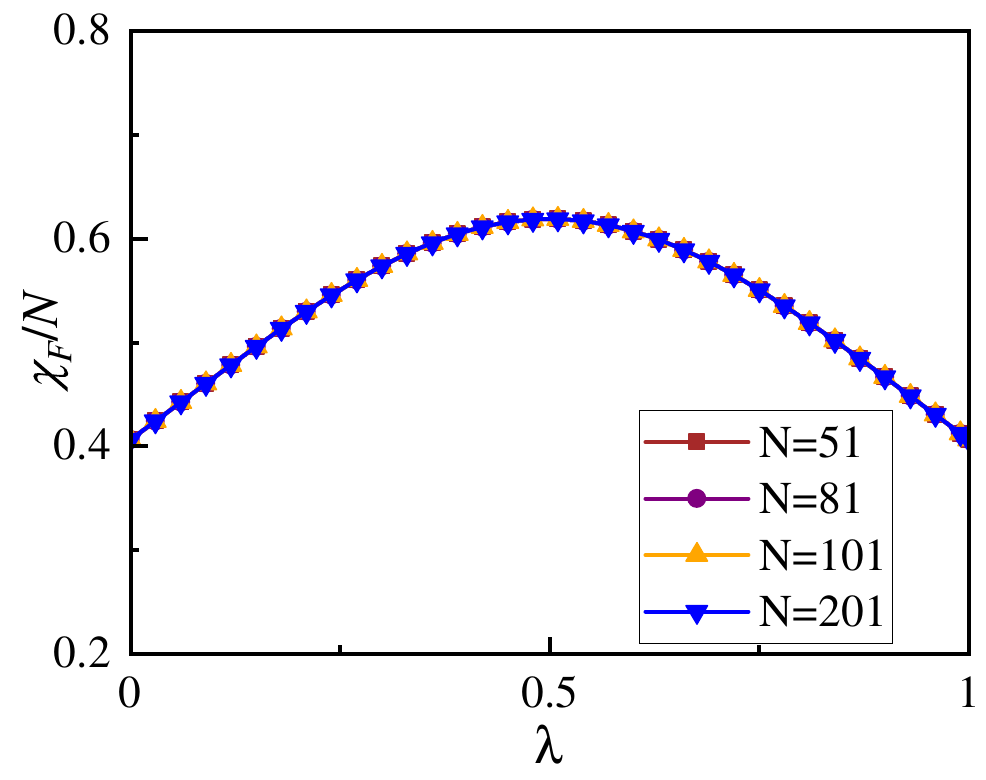}
\caption{(Color online) The finite-size scaling analysis of the fidelity susceptibility per site $\chi_{N}$ for $h=0.5$. The fidelity susceptibility per site does not exhibit a pronounced peak as the system size increases near the transition point.}
\label{sm2}
\end{figure}

In this section, we explore possible QPTs that occur when $h<1.0$. As discussed in the main text, in the two extreme cases of $h=0.0$ and $\lambda=0.0,1.0$, the ground state demonstrates FM long-range order. To investigate the presence of possible quantum phase transitions, we calculated the fidelity susceptibility as a function of $\lambda$ at $h=0.5$, as shown in Fig~\ref{sm2}. The results indicate that the fidelity susceptibility does not exhibit a growing peak with system size. This observation suggests that when $h<1.0$, no phase transition occurs within the system. Consequently, in this regime, the ground state exclusively displays FM long-range order, as illustrated in Fig~\ref{fig1} in the main text.

\section{ADDITIONAL DATA FOR TOPOLOGICAL PHASE TRANSITION}
\label{sec:AppD}
\subsection{Ground state energy density and its second-order derivative}
In this section, we provide additional data on the ground state energy density corresponding to the topological phase transition at $h > 1.0$.

Similar to the main text, the ground state energy of the model is given by Eq.\ref{E5}. Using this equation, we numerically calculate both $\varepsilon_{0}$ and its second-order derivative $-\frac{\partial^{2}\varepsilon_{0}}{\partial \lambda^{2}}$ for $h=2.0,\infty$ at the topological critical point. As depicted in Fig.\ref{sm3}(a) and (b), we observe that the second derivative of the ground state energy density with respect to $\lambda$ becomes sharper at $\lambda=0.5$ as the system size increases, indicating that $\lambda=0.5$ is a continuous QCP.

\subsection{Winding number}

\begin{figure}[t]
\includegraphics[width=0.45\textwidth]{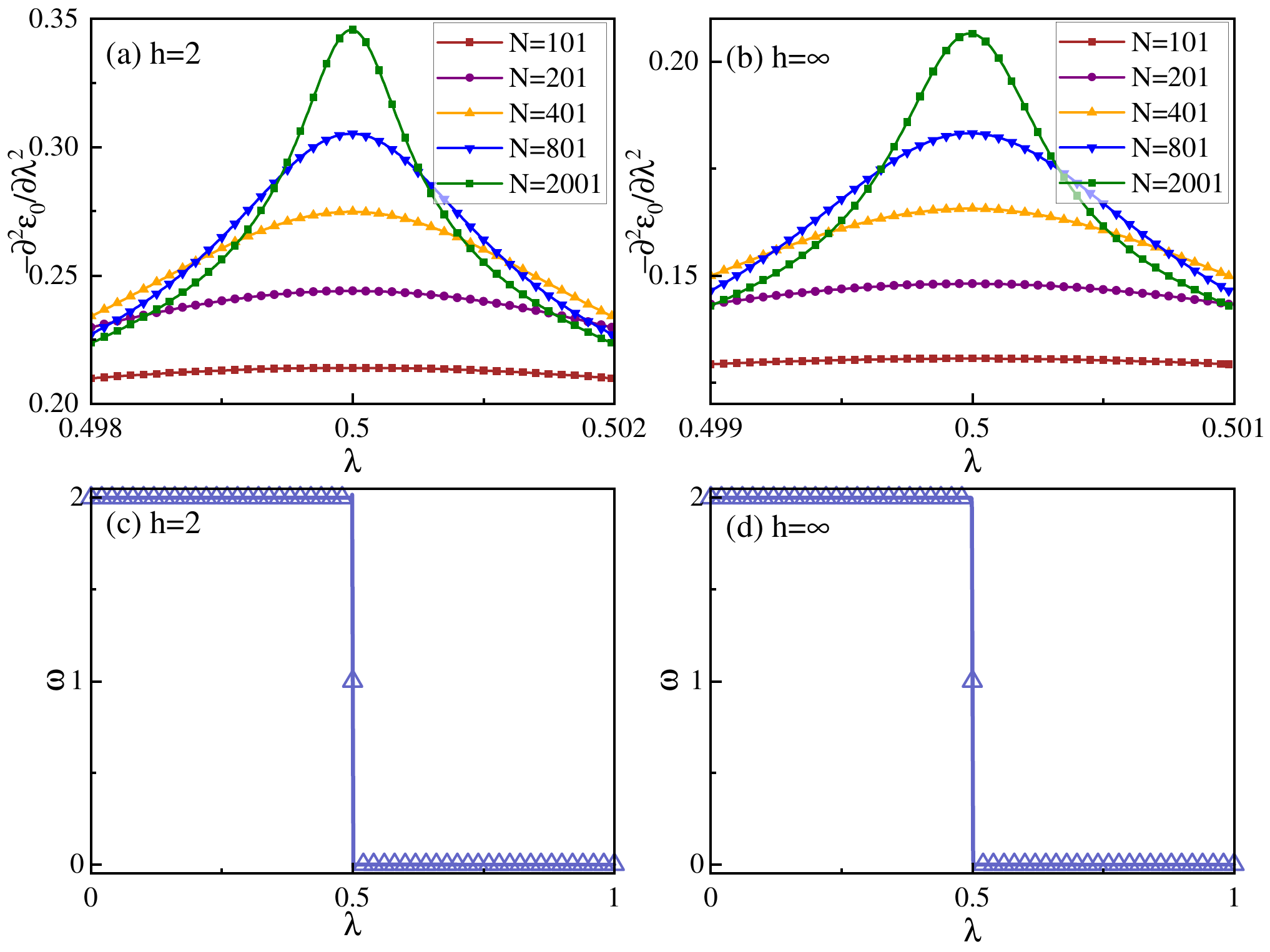}
\caption{(Color online) (a-b) The second derivative of ground-state energy density
$-\frac{\partial^{2}\varepsilon_{0}}{\partial \lambda^{2}}$ with respect to $\lambda$ for $h=2.0,\infty$.
(c-d) The winding number as a function of $\lambda $ with $h=2.0,\infty$ for $N=2000$.}
\label{sm3}
\end{figure} 
In this section, we present additional data on the winding number corresponding to the topological phase transition at $h > 1.0$.

As depicted in Fig.\ref{sm3}(c) and (d), we observe a jump in the winding number at $\lambda=0.5$ for $h=2.0,\infty$, indicating a topological phase transition at this point. Moreover, we note that the winding number takes integer values (0.0 and 2.0) within the trivial PM and SPT phases, respectively, consistent with findings in previous literature\cite{duque2021prb}.

\subsection{Entanglement entropy and central charge}
In this section, we present additional data on the entanglement entropy and central charge corresponding to the topological phase transition at $h > 1.0$.

Similar to the main text, we plot the entanglement entropy $S(l)$ as a function of subsystem sizes $l$ for $h=2.0, \infty$, and $\lambda=0.5$ in the main panel of Fig.\ref{sm4}(c). Additionally, for determining the central charge, we provide the finite-size scaling of entanglement entropy as a function of subsystem sizes ${\rm{log}}l$ for $h = 2.0, \infty$ in the inset of Fig.\ref{sm4}(c). It is evident that the central charge at the conformal critical point consistently equals $1.0$, indicative of its classification within the free boson CFT.

\subsection{Finite-size scaling and critical exponents}
In this section, we present additional data on the finite-size scaling for fidelity susceptibility corresponding to the topological phase transition at $h > 1.0$.

\begin{figure}[t]
\includegraphics[width=0.45\textwidth]{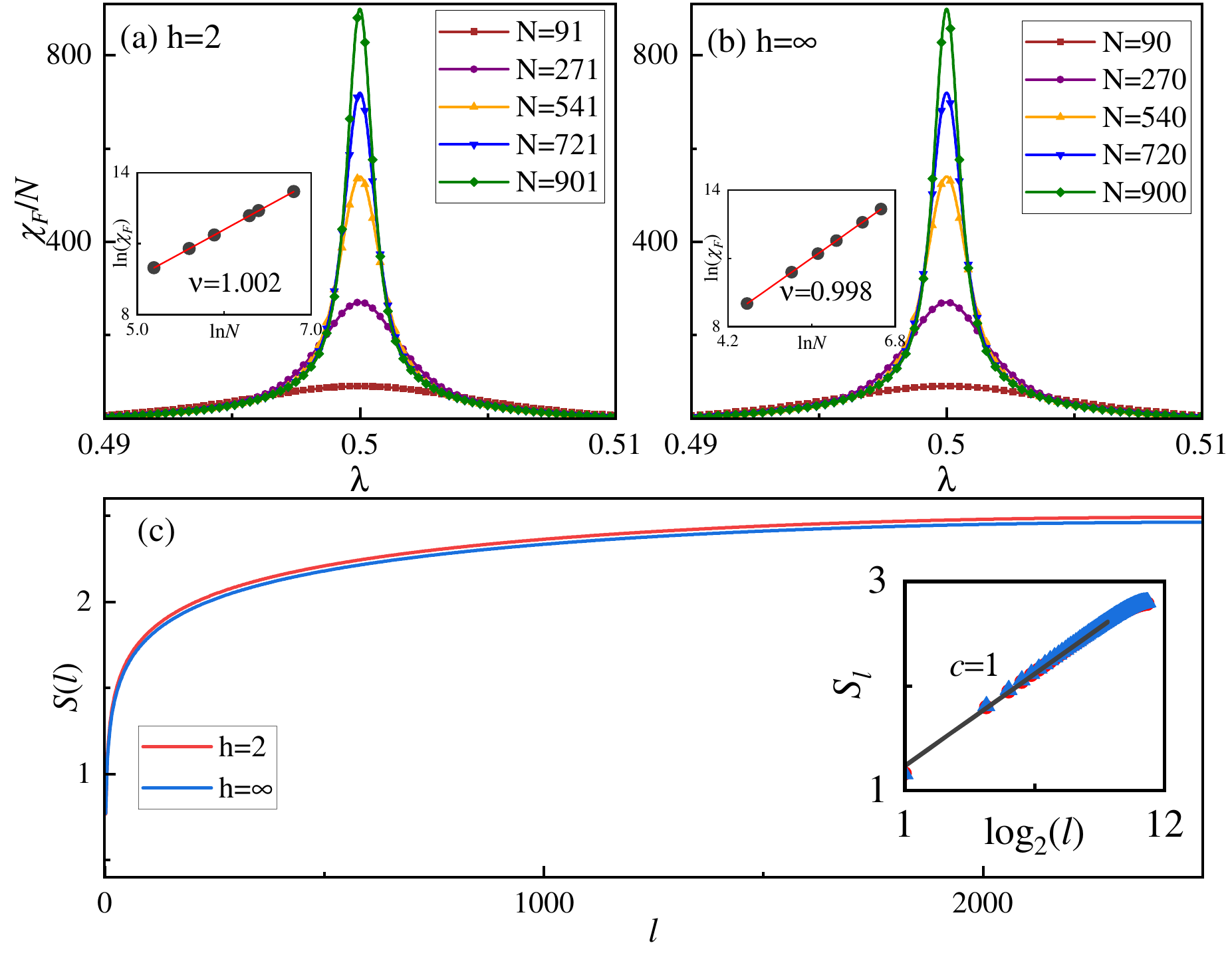}
\caption{(Color online) (a-b) The finite-size scaling analysis of the fidelity susceptibility per site $\chi_{N}$ for $h=2.0$ and $h=\infty$. The fidelity susceptibility per site exhibits a sharp peak near the transition point. The inset displays the log-log plot of the fidelity susceptibility against the system size at the critical point, from which the correlation length critical exponent $\nu=1.0$ can be inferred based on the slope of the fitted straight line. (c) The entanglement entropy as a function of subsystem size $l$ for different $h$ values with $\lambda=0.5$. The central charge $c=1.0$ is obtained through the finite-size scaling fitting of the entanglement entropy (free boson CFT).}
\label{sm4}
\end{figure}

As depicted in Fig.~\ref{sm4}(a) and (b), fidelity susceptibility exhibits a clear peak at $\lambda=0.5$, which becomes sharper as the system size increases, indicating a phase transition occurring at $\lambda=0.5$. According to the scaling relation in Eq.~(\ref{fs_scaling}), we determined the correlation length exponent $\nu=1.0$ for the topological phase transition point for both $h=2.0$ and $\infty$, as shown in the inset of Fig.~\ref{sm4}(a) and (b).

\end{document}